\documentclass[aip, jcp, amsmath, amssymb, reprint, footinbib,floatfix,citeautoscript, longbibliography]{revtex4-1}
\usepackage{tabularx,array}

\usepackage{graphicx}
\usepackage{dcolumn}
\usepackage{bm}
\usepackage{braket}

\usepackage[dvipsnames]{xcolor}
\usepackage[colorlinks=true,
citecolor=purple,
linkcolor=NavyBlue,
urlcolor=NavyBlue]{hyperref}

\usepackage{enumitem}
\usepackage{makecell}
\usepackage{relsize}
\usepackage{booktabs}
\usepackage{siunitx}

\usepackage{todonotes}

\usepackage{upgreek}

\newcommand\T{{\bf T}}

\begin{document}

\author{Anthony J. Dominic III}
\email{Anthony.DominicIII@colorado.edu}
\affiliation{Department of Chemistry, University of Colorado Boulder, Boulder, CO 80309, USA\looseness=-1}

\author{Bertram Cham}
\affiliation{Department of Chemistry, University of Colorado Boulder, Boulder, CO 80309, USA\looseness=-1}

\author{Andr\'{e}s Montoya-Castillo}
\email{Andres.MontoyaCastillo@colorado.edu}
\affiliation{Department of Chemistry, University of Colorado Boulder, Boulder, CO 80309, USA\looseness=-1} 

\title{UGME: A python package for non-Markovian protein folding dynamics}

\date{\today}

\maketitle

\vspace{-0.5in}

\section{Summary}
Biomolecular processes, including protein folding, protein-protein association, and protein-DNA recognition, often occur on timescales beyond what individual atomistic molecular dynamics (MD) simulations can reach. To access these longer-timescale dynamics, practitioners commonly combine many parallelized, short trajectories into reduced kinetic models, such as Markov state models (MSMs) \cite{MSMtext,prinz2011, NoeMSMart18}, that describe how the biomolecules interconvert between a handful of biomolecular configurations (e.g., folded and unfolded states of a protein). Central to MSM construction is the assumption that, at sufficiently long time lags, these projected dynamics become memoryless, such that the future evolution of the system depends only on its current state. Yet, projecting high-dimensional biomolecular dynamics onto a small set of states necessarily discards information about unresolved degrees of freedom, including solvent reorganization, side-chain rearrangements, and intermediate configurations. Because these unresolved motions can continue to influence subsequent dynamics, the resulting dynamics retain memory that is absent from a Markovian description. Here, we present \texttt{ugme}, a modular, open-source Python package for constructing and validating generalized master equations (GME) models that explicitly account for these non-Markovian (memory) effects. 

\begin{figure}[b]
    \includegraphics[width=\linewidth]{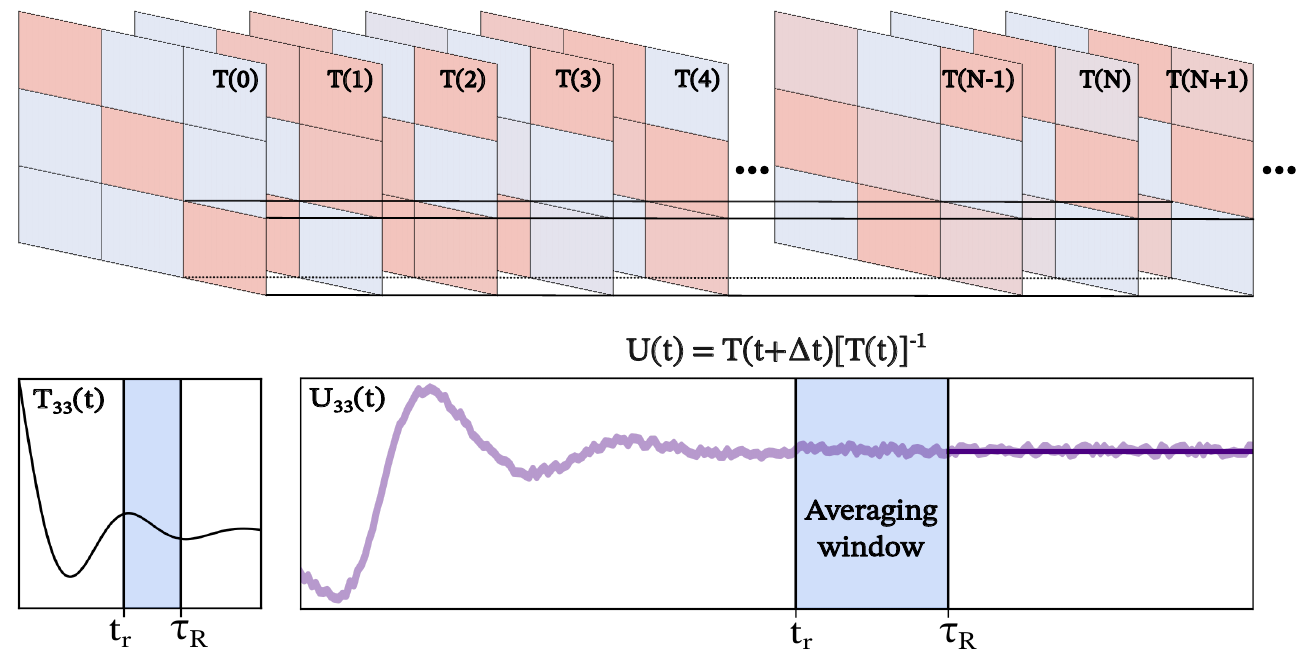}
    \caption{Schematic of the $\mathbf{U}$-GME approach for predicting long-time dynamics from short-time transition probability matrices (TPMs). \textbf{Top:} A time series of 3$\times$3 TPMs, ${\bf T}(t)$, represented as heatmaps, beginning with the identity matrix ${\bf T}(0) = \mathbf{I}$ and extending to the final available time $N$ with timestep $\Delta t$. By applying the \texttt{ugme} software, one can predict the TPM dynamics for times beyond $N$. The black-outlined matrix elements trace the conditional probability of remaining in state 3 shown in the bottom-left panel. \textbf{Bottom left:} Time evolution of ${\bf T}_{33}(t)$, illustrating that the dynamics are still relaxing over the interval defined by the averaging onset $t_r$ and cutoff $\tau_R$. \textbf{Bottom right:} Corresponding evolution of the propagator element ${\bf U}_{33}(t)$ (in light purple), which approaches a stationary value despite the continued evolution of the TPM. The propagator is averaged over the shaded interval $[t_r, \tau_R]$ and subsequently held constant for $t > \tau_R$, enabling prediction of dynamics beyond the available TPM dataset. }
    \label{fig:schematic}
\end{figure}

\section{Statement of need} 
\label{sec:need}

\texttt{ugme} is designed to extend existing molecular kinetics workflows to systems whose projected dynamics retain memory. Current molecular simulation analysis software packages provide well-established tools for trajectory featurization, dimensionality reduction, clustering, and transition probability matrix (TPM) estimation. With a time series of TPMs in hand, existing software packages provide robust methods for estimating, validating, and analyzing MSMs. However, once the dynamics have been projected onto a reduced set of states, accounting explicitly for emergent memory effects generally requires specialized implementations not integrated into standard analysis pipelines. This motivates the development of a tool that can operate directly on TPMs produced by existing workflows.

\texttt{ugme} addresses this need by implementing the noise-resilient time-local GME framework, the $\mathbf{U}$-GME introduced in Ref.~\onlinecite{AJDUGME}. As we show in Fig.~\ref{fig:schematic}, given a time series of TPMs from simulation workflows, analytical models, or experimental sources, the package computes the time-dependent propagator that captures the dynamical memory effects. \texttt{ugme} then identifies when these propagators have converged and uses their long-time behavior to predict dynamics beyond the total time spanned by the input dataset. The package also provides diagnostics for assessing convergence and testing the sensitivity of predicted kinetics to choices such as the memory relaxation time and propagator averaging window. 

Because \texttt{ugme} operates directly on TPMs, it complements existing MD analysis packages such as MSMBuilder, PyEMMA, and deeptime. Users can continue to employ established tools for featurization, clustering, and TPM estimation, while applying \texttt{ugme} alongside MSMs to improve temporal resolution and quantify memory effects from unresolved degrees of freedom, or instead of MSMs when long lag times or underconverged statistics preclude their use.

\section{State of the field}
\label{sec:field}
Established packages for molecular kinetic modeling primarily focus on Markovian descriptions. MSMBuilder,\cite{MSMBuilder} PyEMMA,\cite{pyemma} and deeptime\cite{deeptime} provide workflows for estimating Markov models from simulation data, while msmhelper\cite{msmhelper} provides lightweight tools for estimating, validating, and analyzing MSMs from discrete trajectories. These packages are well-suited to systems for which a sufficiently Markovian state representation and lag time can be identified. However, in Ref.~\cite{AJDUGME}, we demonstrate that a Markovian description is insufficient for capturing mRNA recognition dynamics by the human Argonaute complex.

GMEs extend this framework by retaining the influence of unresolved degrees of freedom on the projected dynamics. Memory kernel formulations encode this influence through a time-convolution integral, whereas time-local formulations describe it through an explicitly time-dependent propagator or generator, akin to a time-dependent rate matrix. While the two representations are formally similar, their numerical implementations differ substantially. \texttt{ugme} implements a time-local construction that requires only a sequence of TPMs. This formulation avoids direct evaluation of projected forces, numerical differentiation of correlation functions, or explicit convolution of a memory kernel.  

The scope of \texttt{ugme} therefore differs from existing MSM packages. While existing tools provide machinery for constructing and analyzing Markovian kinetic models, \texttt{ugme} begins from the resulting TPMs and analyzes the memory retained in their dynamics. Keeping these steps separate allows \texttt{ugme} to remain compatible with TPMs produced by existing software while providing a workflow centered specifically on the estimation, validation, and propagation of non-Markovian kinetic models.

\section{Software design}
\label{sec:design}

The primary input to \texttt{ugme} is a time-indexed sequence of square TPMs, $\left\{\T(t_0),\T(t_1),\ldots,\T(t_N)\right\}$, where each $\T(t_n)$ contains the conditional transition probabilities between the states of the reduced model at time $t_n = n \Delta t$, where $\Delta t$ is the time interval between successive TPMs (i.e., the temporal resolution of the reference dataset) and $t_N$ is the final timestep of the input TPM dataset. Crucially, their time evolution evolves according to $\T(t_{n+1}) = \mathbf{U}(t_n) \T(t_n)$, where the time-dependent propagator $\mathbf{U}(t_n)$ provides a direct measure of the relaxation of memory in the reduced dynamics. The memory in $\mathbf{U}(t_n)$ decays and non-Markovian effects become unimportant when its matrix elements reach a steady state. As such, the memory relaxation timescale $\tau_R \equiv t_R = R \Delta t$ is the time at which $\mathbf{U}(t_n)$ plateaus. For times beyond $\tau_R$, we replace the propagator with its stabilized value $\mathcal{U}_\infty (\Delta t)  \equiv \mathcal{U}(\tau_R)$. Beyond $\tau_R$, the dynamics evolve according to 
\begin{equation}
\label{eq:ugmeforecase}
    \T(\tau_R + n\Delta t) = \mathbf{U}_\infty^n (\Delta t) \T(\tau_R).
\end{equation}


The package is organized around the \texttt{UGME} model object, which stores the TPM time series, temporal resolution, and quantities generated during model construction. From the TPMs, \texttt{ugme} computes the corresponding time-local propagator $\mathbf{U}$ by directly inverting Eq.~\eqref{eq:ugmeforecase} and tracks its time evolution. The plateauing of $\mathbf{U}(t)$ quantifies the memory's relaxation time in the reduced dynamics. For sufficiently converged data, \texttt{ugme} identifies the memory-relaxation timescale $\tau_R$ using the RMSE-based convergence criterion we established in Ref.~\onlinecite{AJDUGME}. 

TPMs estimated from finitely-sampled trajectory data can introduce statistical error into the late-time propagator. To reduce the propagator's sensitivity to this noise, \texttt{ugme} supports time-averaging of the propagator $\mathbf{U}$. We introduce tools to compute the onset ($t_r$) and offset ($\tau_R$) averaging parameters and the averaged propagator that we define as
\begin{equation}
\label{eq:avgprop}
    \langle \mathbf{U}_\infty(\Delta t) \rangle  = \frac{1}{R - r}\sum_{n = r}^{R-1} \mathbf{U}(t_n).
\end{equation}
From Eq.~\eqref{eq:avgprop}, we suppress the noise and propagate the dynamics as 
\begin{equation}
\label{eq:nonoise}
    \T(\tau_R + n\Delta t) = \langle \mathbf{U}_\infty(\Delta t) \rangle ^n \T(\tau_R).
\end{equation}

In a typical workflow, a user supplies a TPM time series, constructs the corresponding \texttt{UGME} model, determines $\tau_R$, selects an averaging window when needed, and propagates to access the long-time dynamics. This organization keeps the model diagnostics accessible while maintaining a compact interface for routine calculations. These stages are exposed separately so that users can inspect the quantities entering the model and assess if and how numerical choices affect the resulting kinetics. 

The repository includes source code, API documentation, example NumPy datasets, and a notebook demonstrating TPM loading, propagator construction, reconstruction, relaxation-time scans, averaging-window scans, and long-time prediction. These materials provide a reproducible starting point for applying the method to new biomolecular systems and for extending the implementation to additional GME estimators.

\section{Research impact statement}
\label{sec:impact}

We developed the methods implemented in \texttt{ugme} to extract long-time biomolecular kinetics from short-time simulation data. In Ref.~\onlinecite{AJDUGME}, we applied the $\mathbf{U}$-GME to alanine dipeptide, the FiP35 WW Domain, and the human Argonaute complex, showing that the method can recover long-time relaxation dynamics from substantially shorter input trajectories than required by the conventional MSM treatment. We subsequently discussed the broader role of memory-enriched models for biomolecular simulations in Ref.~\onlinecite{AJD-perspective-2023}

Since then, researchers have adopted their own time-local formulation. For example, Ref.~\onlinecite{Liu2026OptimalData} has adapted the time-local GME formulation to estimate discrete time-local GMEs from experimental and simulation data by constructing physically constrained maximum likelihood estimators and applied these methods to experimental datasets corresponding to single-molecule F\"orster resonance energy transfer (FRET) in an ion channel, experimental nanoparticle tracking of a processive molecular motor, and the simulated protein folding dynamics of HP35. Beyond biophysics, the time-local GME has extended to polymer dynamics,\cite{Bement2025ModelsTechniques, Sinclair2025InfluenceMigration} transport,\cite{Bhattacharyya2025NonequilibriumDiffusion, Bhattacharyya2025Space-localSimulation} and spectroscopy.\cite{Sayer2026Short-livedEvolution} Collectively, these studies illustrate that short-lived memory can be exploited across a range of dynamical problems where projecting onto a reduced set of variables introduces memory effects. 

By separating GME analysis from TPM estimation, \texttt{ugme} makes these explicit approaches to capturing memory effects readily available for analyzing molecular dynamics. We anticipate that this interface will facilitate applications of time-local GME models to new biomolecular systems while also providing a foundation for testing and extending memory-enriched kinetic modeling methods.

\section{AI usage disclosure}
\label{sec:AI}

A significant portion of this code was developed from 2021-2023 without the use of AI tools. AI tools were used to proofread code, proofread the publication, and build initial docstrings and comments for portions of the code that previous implementations of this code did not have. 

\section{Data Availability}
Our materials are freely available at the following GitHub repository: \url{https://github.com/ajdominic/ugme.github.io/tree/main}.


\section{Acknowledgments}
\label{sec:acknowledgements}

A.J.D., B.C., and A.M.C.~acknowledge support by an Early Career Award in the CPIMS program in the Chemical Sciences, Geosciences, and Biosciences Division of the Office of Basic Energy Sciences of the U.S.~Department of Energy under Award DE-SC0024154. A.J.D.~also acknowledges partial support from the NIH/CU Molecular Biophysics Program and the NIH Biophysics Training Grant T32 GM145437. This work was supported by an Alfred P. Sloan Research Fellowship and the Camille and Henry Dreyfus Foundation through a Camille Dreyfus Teacher-Scholar Award. This work utilized the Alpine high performance computing resource at the University of Colorado Boulder. Alpine is jointly funded by the University of Colorado Boulder, the University of Colorado Anschutz, Colorado State University, and the National Science Foundation (Award No. 2201538).


\section{References}
\bibliography{references.bib}

\end{document}